\documentclass[3p,times]{elsarticle}

\usepackage{ecrc}

\volume{00}

\firstpage{1}

\journalname{Nuclear Physics A}

\runauth{J.~Jia}

\jid{nupha}

\jnltitlelogo{Nuclear Physics A}

\usepackage{graphicx}
\usepackage{amsmath,amssymb}
\newcommand{\pp}{\mbox{$p$+$p$}}
\newcommand{\pPb}{\mbox{$p$+Pb}}

\newcommand{\pta}{\mbox{$p_{\mathrm{T}}^{\mathrm{a}}$}}
\newcommand{\ptb}{\mbox{$p_{\mathrm{T}}^{\mathrm{b}}$}}

\newcommand{\pA}{\mbox{$p$+A}}
\newcommand{\PbPb}{\mbox{Pb+Pb}}

\newcommand{\nch}{\mbox{$N_{\mathrm{ch}}$}}

\newcommand{\npart}{\mbox{$N_{\mathrm{part}}$}}

\newcommand{\sqrtsnn}{\mbox{$\sqrt{s_{\mathrm{NN}}}$}}
\newcommand{\Dphi}{\mbox{$\Delta \phi$}}
\newcommand{\Deta}{\mbox{$\Delta \eta$}}

\newcommand{\pT}{\mbox{$p_{{\mathrm{T}}}$}}
\newcommand{\eT}{\mbox{$E_{{\mathrm{T}}}$}}

\newcommand{\npartf}{N_{\mathrm {part}}^{\mathrm{F}}}
\newcommand{\npartb}{N_{\mathrm {part}}^{\mathrm{B}}}
\begin{document}

\begin{frontmatter}

%% Title, authors and addresses

%% use the tnoteref command within \title for footnotes;
%% use the tnotetext command for the associated footnote;
%% use the fnref command within \author or \address for footnotes;
%% use the fntext command for the associated footnote;
%% use the corref command within \author for corresponding author footnotes;
%% use the cortext command for the associated footnote;
%% use the ead command for the email address,
%% and the form \ead[url] for the home page:
%%
%% \title{Title\tnoteref{label1}}
%% \tnotetext[label1]{}
%% \author{Name\corref{cor1}\fnref{label2}}
%% \ead{email address}
%% \ead[url]{home page}
%% \fntext[label2]{}
%% \cortext[cor1]{}
%% \address{Address\fnref{label3}}
%% \fntext[label3]{}

%\title{Correlations and fluctuations in high-energy nuclear collisions - a ``flow''-centric review}
\title{Collective phenomena in high-energy nuclear collisions}

%% Single author (and collaboration) - please insert
%\author{Author (for the XYZ\fnref{col1} Collaboration)}
%\fntext[col1] {A list of members of the XYZ Collaboration and acknowledgements can be found at the end of this issue.}
\author{Jiangyong Jia}
\address{Department of Chemistry, Stony Brook University, Stony Brook, NY 11794, USA,}
\address{Physics Department, Brookhaven National Laboratory, Upton, NY 11796, USA}

%% For multiple authors, replace the above by:

%\author[label1]{Author1}
%\author[label2]{Author2}

%\address[label1]{Address 1}
%\address[label2]{Address 2}

\begin{abstract}
%\verb!http://www.elsevier.com/author-schemas/preparing-crc-journal-articles-with-latex!.
I review experimental studies of collective phenomena in $\pA$ and A+A collisions presented in the Quark Matter 2014 conference.
\end{abstract}

\begin{keyword}
%% keywords here, in the form: keyword \sep keyword
Heavy-ion collisions \sep Fluctuations \sep Correlations \sep Collective flow \sep Ridge
%% MSC codes here, in the form: \MSC code \sep code
%% or \MSC[2008] code \sep code (2000 is the default)

\end{keyword}

\end{frontmatter}
%%
%% Start line numbering here if you want
%%
% \linenumbers

%% main text

\section{Introduction}
\label{intro}
%contributed significantly to our present unders

%various correlation observables, which are sensitive to the fluctuations generated at various stage of the space-time evolution of the system. One primary example is the collective flow, 
%, manifested as azimuthal anisotropy of particle production in the transverse plane.
%The collective flow, one of the hallmark of A+A collisions, in combination with viscous hydrodynamic models has been successfully used to infer the initial condition and transport properties of the matter. Recently, qualitatively similar collective behaviour has also been observed in small collision system such as high-multiplicity p+p and $\pA$ collisions. The relevant question is whether he
% Clearly the apparent collective behaviours are manifestation of QCD in different high-density systems. But re

The goal of high-energy nuclear experiments is to understand the properties of the QCD matter created in the heavy-ion collisions. Most progress to this end has been obtained from study of collective flow phenomena, which are sensitive to the dynamics of the system at various stages of the space-time evolution. Extensive measurements of various flow observables in A+A collisions at RHIC and the LHC, in combination with successful modeling by relativistic viscous hydrodynamics, have placed important constraints the transport properties and initial conditions of the produced matter~\cite{Heinz:2013th,Luzum:2013yya}.

Recently, qualitatively similar collective behaviour has also been observed in high-multiplicity $\pA$ collisions~\cite{CMS:2012qk}. Current debate is focused on the effective mechanism behind the apparent collectivity in these small systems: Is it initial state effect, final state effect or both? 
%This conference has seen new measurements that can help address this question.
%The came as a surprise, since the transverse size of the produced system was thought to be too small for the hydrodynamic flow description to be applicable. 

This proceedings reviews the new results from $\pA$ and A+A collisions from RHIC and the LHC, obtained with various flow observables. I first summarize experimental progresses in studying the collectivity in small collision systems. I then make a few remarks on flow results in A+A collisions that I personally found interesting. This is followed by a focused discussion of the event-by-event (EbyE) flow fluctuations measurements in A+A collisions. Finally I discuss open issues and future physics opportunities.

\vspace*{-0.2cm}
\section{Collectivity in small system}
The observation of collective behaviour in high-multiplicity $\pA$ collisions~\cite{CMS:2012qk} came as a surprise, since the transverse size of the produced system was thought to be too small for the hydrodynamic description to work. Experimentalists attempt to elucidate this puzzle by mapping out the detailed properties of the azimuthal harmonics in $\pA$ collisions and comparing with those in peripheral A+A collisions.
% on detailed properties of the flow harmonics in $\pA$ collisions and compare with those in peripheral A+A collisions. Experimentalist resolving this puzzle by mapping out the detailed properties of the flow harmonics in $\pA$ collisions and compare with those in peripheral A+A collisions. 

\begin{figure}
\centering
\vspace*{-0.1cm}
\includegraphics[width=0.9\linewidth]{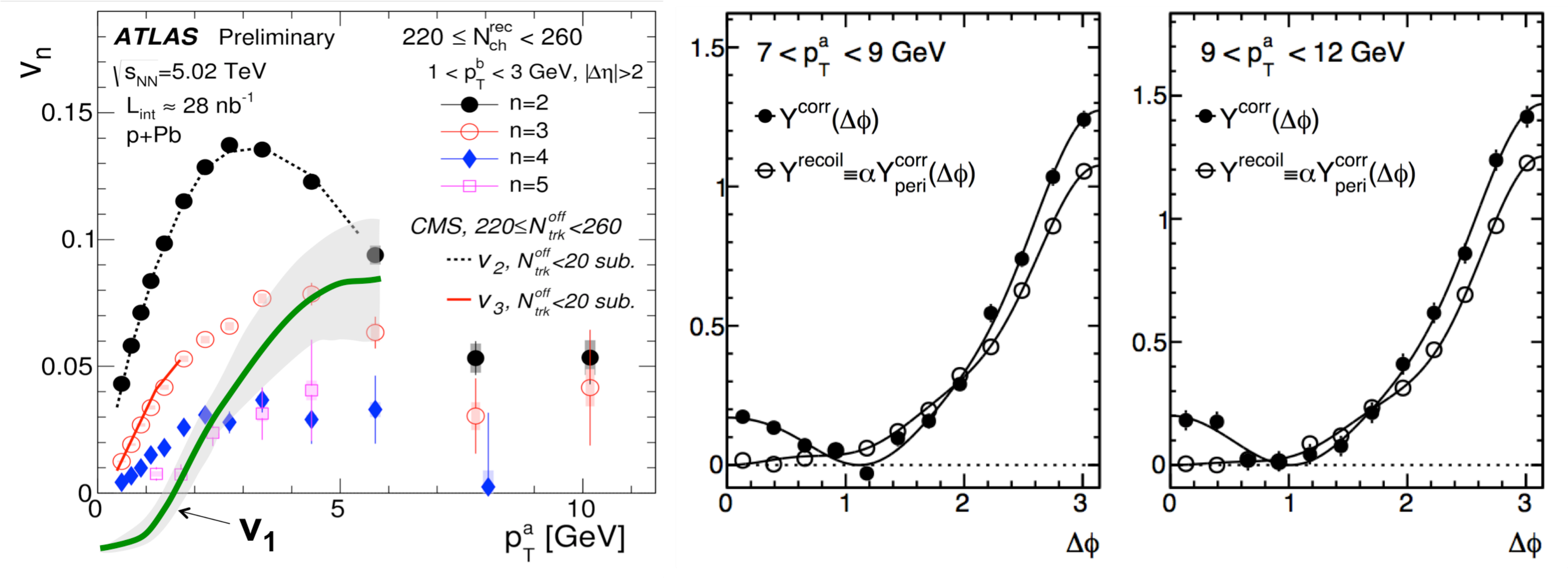}
\vspace*{-0.3cm}
\caption{\label{fig:1} The $v_n(\pT)$ with $n=2$ to 5 for pairs with $|\Deta|>2$~\cite{sooraj}. An overlay sketch of rapidity-even $v_1$ data is also shown. Results are compared to the CMS data with comparable multiplicity.}
\end{figure}

Figure~\ref{fig:1} (a) shows the first five azimuthal harmonics $v_1$ to $v_5$ as a function of $\pT$ in high-multiplicity $\pPb$ collisions, obtained from a two-particle correlation (2PC) method~\cite{sooraj}. The away-side and near-side short-range correlations have been estimated from events with low multiplicity and subtracted. The $v_n$ magnitude is largest for $n=2$, and decreases with increasing $n$. All $v_n$ increase with $\pT$ up to 3--5 GeV and then decrease, but remain positive at higher $\pT$. It is interesting that the $v_2$ and $v_3$ values are sizable at $\pT>8$ GeV (3-5\%), reflecting a significant near-side ridge regardless of the recoil subtraction procedure as shown in the right panels of Figure~\ref{fig:1}. This maybe the first indication of a small azimuthal anisotropy of jet yield in high-multiplicity $\pPb$ collisions.

Figure~\ref{fig:1} also shows an overlay sketch of the preliminary rapidity-even $v_1$~\cite{sooraj}. The values of $v_1$ are comparable to $v_3$ at $\pT\sim4$ GeV, albeit with larger uncertainties since they are directly sensitive to the recoil subtraction. 
\begin{figure}[!h]
\centering
\vspace*{-0.1cm}
\includegraphics[width=0.68\linewidth]{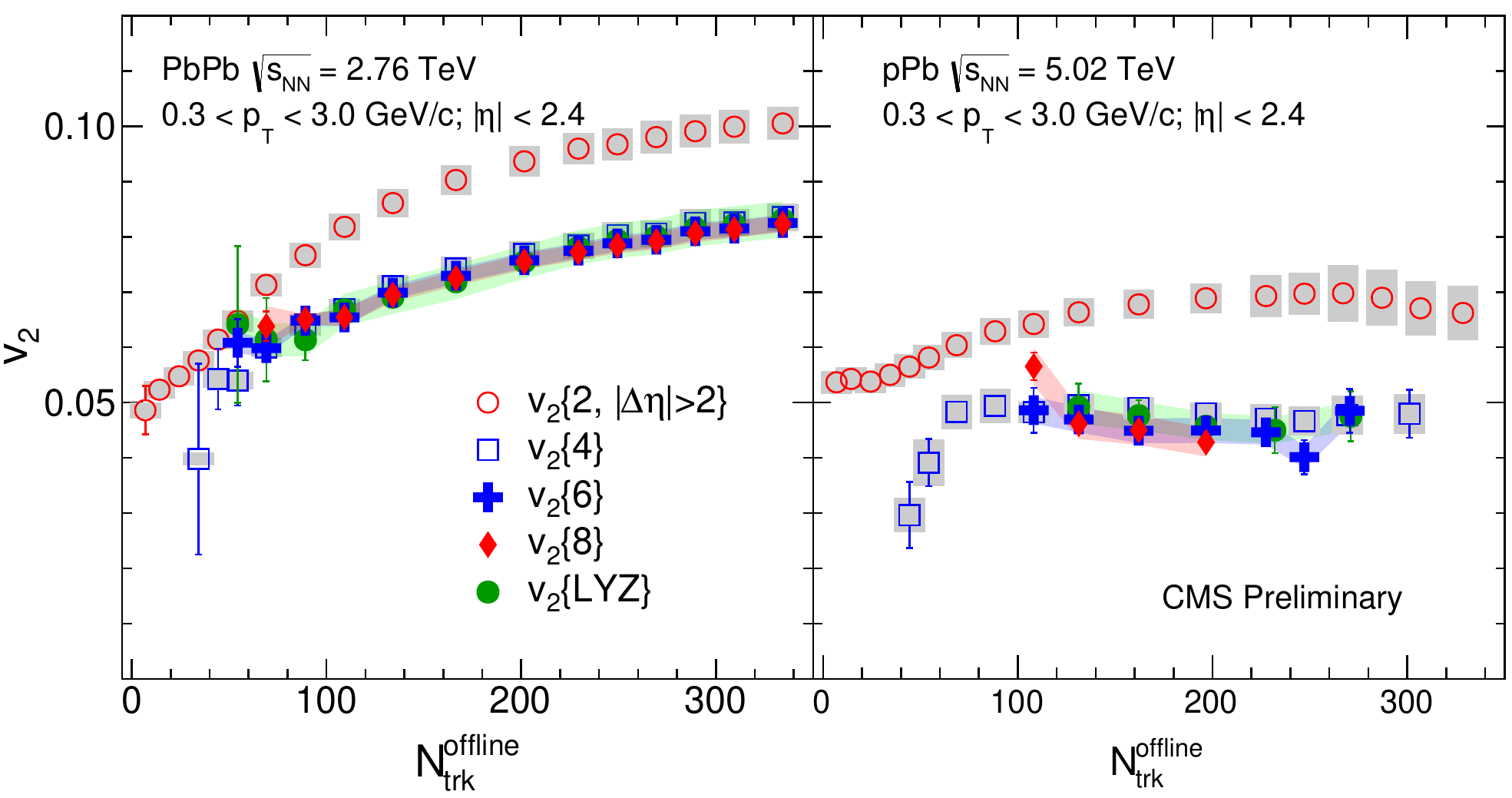}
\vspace*{-0.3cm}
\caption{\label{fig:2} The $v_2$ results obtained from two, four, six, eight-particle cumulants, as well as Lee-Yang Zero (LYZ) method, averaged over the $\pT$ range of 0.3--3.0 GeV, as a function of number of charged particles in $\PbPb$ (left panel) and $\pPb$ (right panel) collisions~\cite{quanwang}.}
\end{figure}

Multi-particle cumulants beyond 2PC have been argued to be able to distinguish the correlation arising from collective dynamics from those arise from sources involving a few particles. CMS presents a detailed measurement of multi-particle cumulants $v_2\{4\}$, $v_2\{6\}$ and $v_2\{8\}$, as well as  Lee-Yang Zero (LYZ) method $v_2\{\mathrm{LYZ}\}$~\cite{quanwang}. They agree within $\pm10$\% over a broad multiplicity range in $\pPb$ or $\PbPb$ collisions (Figure~\ref{fig:2}), confirming the collective nature of the observed correlations in both systems. The apparent difference from $v_{2}$\{2\} reflects the fluctuations either present in the initial geometry or generated in the collective expansion. 

\begin{figure}[!h]
\centering
\vspace*{-0.1cm}
\includegraphics[width=0.73\linewidth]{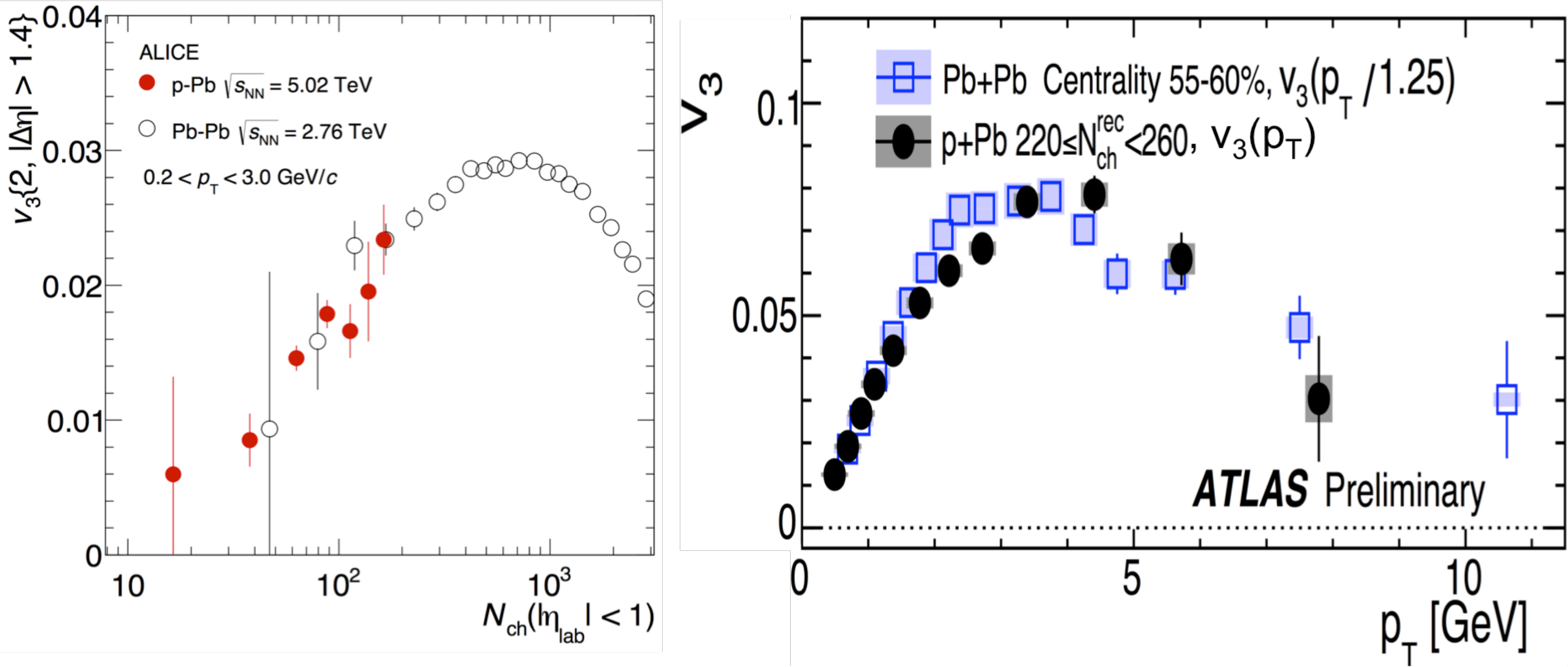}
\vspace*{-0.3cm}
\caption{\label{fig:3}  The $v_3$ data compared between $\pPb$ and $\PbPb$ collisions as a function of charged particle multiplicity in $|\eta|<1$~\cite{timmons} (left panel) and as a function of $\pT$~\cite{sooraj} at approximately the same multiplicity (right panel) in the two systems. In the right panel, the $\PbPb$ data are rescaled horizontally by a constant factor of 1.25 to match the $\langle\pT\rangle$ in $\pPb$ collisions.}
\end{figure}

Hydrodynamic is the commonly accepted description for $v_n$ in central and mid-central A+A collisions. However, as the system size decreases, this description is expected to break down at some point. Comparison of $v_n$ between $\pPb$ and $\PbPb$ at similar multiplicity can help us to understand where and how the break down takes place. Figure~\ref{fig:3} compares the multiplicity dependence of the integral $v_3$ (left panel) and the $\pT$ dependence of $v_3$ (right panel)~\cite{timmons} between the two collision systems. Very similar multiplicity and $\pT$ dependence are observed between the two systems, suggesting that the collectivity is controlled by total multiplicity, as expected from a argument based on conformal hydrodynamic~\cite{Basar:2013hea}. This conformal scaling argument also qualitatively explains the similarity of the multiplicity dependence of the HBT radii in $\pp$, $\pPb$ and $\PbPb$ collisions~\cite{hbt}. CMS also presents a precision measurement of $v_2$ and $v_3$ for $K_{s}^0$ and $\Lambda$ in $\pPb$ collisions~\cite{sharma}, a clear particle-mass dependence is seen that is persist to at least 5 GeV in $\pT$ with a magnitude similar to values in $\PbPb$ collisions at comparable multiplicity. 

\begin{figure}[ih]
\centering
\vspace*{-0.1cm}
\includegraphics[width=0.76\linewidth]{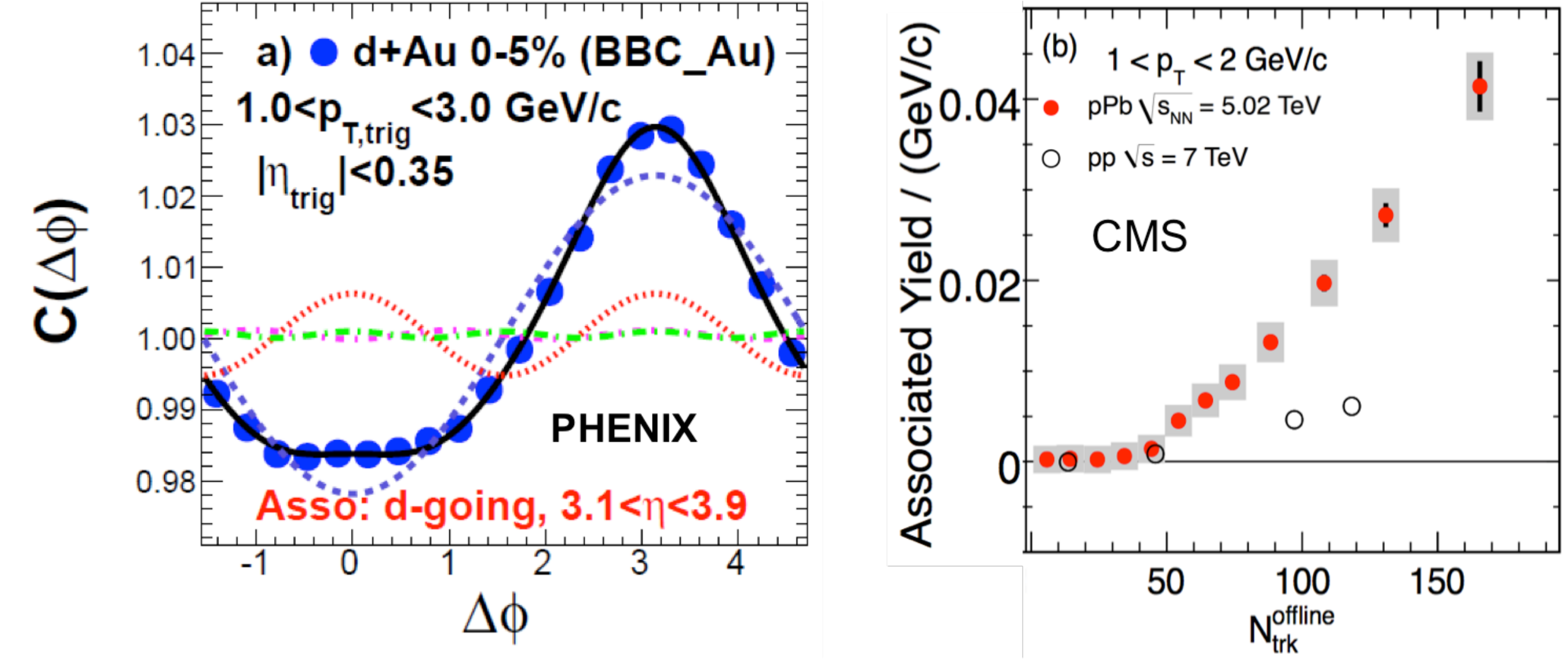}
\vspace*{-0.3cm}
\caption{\label{fig:0} (a) Azimuthal correlations between mid-rapidity hadrons and forward $\eT$ in the d-going direction in $\sqrtsnn=200$ GeV d+Au collisions~\cite{shengli}. (b) The per-trigger yield for particle pairs selected in $1<\pT<2$ GeV as a function of charged particle multiplicity in $\sqrtsnn=2.76$ TeV $\pPb$ collisions~\cite{CMS:2012qk}.}
\end{figure}

Here I would like to caution people about one technical aspect of those 2PC analyses that rely on the ZYAM (zero-yield at minimum) procedure. In this procedure, the near-side excess associated with the long-range ridge is obtained by subtracting a constant pedestal that is matched to the minimum of the correlation function. Due to the large and broad away-side jet peak (especially at low $\pT$), the tail of the away-side peak tends to feed into the near-side region and create a minimum at $\Dphi_{\mathrm{ZYAM}}=0$. As long as the ridge yield is not enough to fill the near-side jet valley, one always finds zero ridge yield despite the apparent change in the near-side shape of the correlation function (see Figure~\ref{fig:0}(a) from~\cite{shengli}). Consequently, the multiplicity dependence of the extracted near-side per-trigger yield exhibits a step-like structure at small $\nch$~\cite{CMS:2012qk} (Figure~\ref{fig:0}(b)). This step could be misinterpreted as the onset of ridge or saturation physics.

\vspace*{-0.3cm}
\section{Comments on selected flow results in A+A collisions}
Several new flow results are worthy of mentioning. ALICE publishes a detailed measurement of $v_2$ for many identified particles~\cite{Abelev:2014pua}. Comparing to the RHIC results, ALICE results imply stronger radial flow and hadronic re-scattering effects. The latter is responsible for a poor number-of-constituent-quark (NCQ) scaling for different particle species. In particular, the NCQ-scaling of $\phi$-meson $v_2$ suggests that it flows like a baryon in central collisions and like a meson in mid-central collisions. This could happen if $\phi$ meson undergoes significant hadronic re-scattering in central collision where the radial flow is the largest, despite its small hadronic cross-section.

New flow results in Cu+Au and U+U collisions are reported by PHENIX and STAR~\cite{cuauuu}. Significant $v_1$ is measured in Cu+Au collisions with respect to the spectators, reflecting the average dipolar geometry in these collisions. The dependence of the multiplicity dependence of $v_2$ in ultra-central U+U collisions is found to differ from that in the Au+Au collisions, reflecting the influence of the deformed geometry in U+U collisions. However, each collision system introduces its own uncertain in collision geometry, which need to be understood in order to make good use of these new results.

Due to event-by-event fluctuations, it was predicted that the event-plane angle $\Phi_n$ fluctuates in $\pT$, leading to a breakdown of the factorization in extracting the $v_n$ from azimuthal harmonics $V_{n\Delta}$ in two-particle correlations~\cite{Gardim:2012im}. This breakdown can be quantified by the $r_n$ factor:
\begin{eqnarray}
r_n = \frac{V_{n\Delta}(\pta,\ptb)}{\sqrt{V_{n\Delta}(\pta,\pta)V_{n\Delta}(\ptb,\ptb)}}\approx\frac{\left\langle v_n(\pta)v_n(\ptb)\cos n(\Phi_n(\pta)-\Phi(\ptb))\right\rangle}{\sqrt{\left\langle v_n^2(\pta)\right\rangle\left\langle v_n^2(\ptb)\right\rangle}}
\end{eqnarray}
$r_n$ is expected to be less than one if $\Phi_n$ depends on $\pT$. Experimental data show significant factorization breaking for $v_2$ in ultra-central $\PbPb$ collisions on the order of up to 20\% ($r_2$ reaches 0.8), consistent with viscous hydrodynamic model predictions~\cite{CMS:2013bza}. However, the breaking is much smaller for $v_2$ in other centrality ranges and for other harmonics in the full centrality range~\cite{CMS:2013bza,ATLAS:2012at}. New data in $\pPb$ collisions from CMS and ALICE suggest the level of breaking is at most a few percents in low $\pT$ region~\cite{fac}. But even some of these deviations could be due to nonflow effects from jets, hence the use of recoil subtraction procedure~\cite{sooraj} could be important.

%~\cite{Qiu:2011iv}
The $v_n$ spectrum in ultra-central A+A collisions post a challenge for theory. The original motivation is that the initial conditions in these collisions are predominantly generated by fluctuations such that the magnitudes of first several $\varepsilon_n$ are comparable, and that the hydrodynamic response is expected to be linear $v_n\propto \varepsilon_n$ for all harmonics. Hence these collisions are expected to provide better constraints on the mechanism of the density fluctuations. The precision data from ATLAS~\cite{ATLAS:2012at} and CMS~\cite{CMS:2013bza} show several features that are not described by models: 1) $v_3$ is comparable or larger than $v_2$, which is challenging since naively one expect $\varepsilon_3\approx \varepsilon_2$, but the $v_3$ should suffer larger viscous correction. 2) The $v_2(\pT)$ has very different shape comparing to other harmonics; It peaks at much lower $\pT$, around 1.5 GeV compare to 3-4 GeV for other harmonics. Calculation in Ref.~~\cite{denicol} shows that the nucleon-nucleon correlation and effects of bulk viscosity can reduce the $v_2$ values relative to $v_3$.
% 3) The $v_2$ is also observed to break the factorization relation at 20-30\% level, while such breaking is much less in other centrality and for higher-order harmonics in all centrality~\cite{Aad:2012bu,CMS:2013bza}. The current hydro calculations describe the factorization data for $v_2$ but over-predict those for the $v_3$~\cite{Heinz:2013bua}. 

\vspace*{-0.3cm}
\section{Event-by-event flow fluctuations in A+A collisions}
In heavy-ion collisions, the number of participating nucleons $\npart$ is finite and their positions fluctuate randomly in the transverse plane, leading to strong EbyE fluctuations of $\varepsilon_n$ and participant plane angle $\Phi_n^*$. Consequently, the matter created in each collision has it own set of flow harmonics, described by a joint probability distribution (pdf):
\begin{equation}
\label{eq:220}
p(v_n,v_m,...., \Phi_n, \Phi_m, ....)=\frac{1}{N_{\mathrm{evts}}}\frac{dN_{\mathrm{evts}}}{dv_ndv_m...d\Phi_{n}d\Phi_{m}...}.
\end{equation}
Some projections of this pdf, such as $p(v_n)$, can be measured directly via unfolding method~\cite{Aad:2013xma}. But they are more generally studied via $m$-particle correlations of the following form~\cite{Bhalerao:2011yg}:
\begin{eqnarray}
\label{eq:223}
\left\langle \cos(n_1\phi_1+n_2\phi_2...+n_m\phi_m)\right\rangle=\left\langle v_{n_1}v_{n_2}...v_{n_m} \cos(n_1\Phi_{n_1}+n_2\Phi_{n_2}...+n_m\Phi_{n_m})\right\rangle, \Sigma n_i=0.
\end{eqnarray}
For example (more can be found in Ref.~\cite{Jia:2014jca}), the two-particle cumulant of $p(v_n)$ is obtained as $\left\langle \cos (n\phi_1-n\phi_2)\right\rangle = \left\langle v_n^2\cos (n\Phi_n-n\Phi_n) \right\rangle = \left\langle v_n^2 \right\rangle$; the lowest-order cumulant for $p(v_n,v_m)$ requires a four-particle correlation of the form~\cite{Bilandzic:2013kga}:
\small{\begin{eqnarray}
\left\langle \cos (n\phi_1-n\phi_2+m\phi_3-m\phi_4)\right\rangle_c = \left\langle \cos (n\phi_1-n\phi_2+m\phi_3-m\phi_4)\right\rangle -\left\langle \cos (n\phi_1-n\phi_2)\right\rangle\left\langle \cos (m\phi_3-m\phi_4)\right\rangle = \left\langle v_n^2 v_m^2 \right\rangle-\left\langle v_n^2 \right\rangle\left\langle v_m^2 \right\rangle\;;
\end{eqnarray}}\normalsize
the event-plane correlations $p(\Phi_n,\Phi_m,...)$ can be accessed via a $\Sigma c_i$-particle correlation of the form:
\begin{eqnarray}
\langle \cos (\Sigma_{i_1=1}^{c_1}\phi_{i_1}+\Sigma_{i_2=1}^{c_2}2\phi_{i_2}+...+\Sigma_{i_l=1}^{c_l}l\phi_{i_l})\rangle=
\left\langle v_1^{c_1}v_2^{c_2}..v_l^{c_l}\cos(c_1\Phi_1+2c_2\Phi_2...+lc_l\Phi_l)\right\rangle,\;\;\;\Sigma ic_i=0.
\end{eqnarray}

In the last few years, impressive progresses have been made in studying these flow observables~\cite{Jia:2014jca}. We now understand that $v_2$ and $v_3$ are the dominant harmonics, driven mainly by the linear response to the ellipticity and triangularity of the initially produced fireball~\cite{Qiu:2011iv}: $v_2e^{i2\Phi_2} \propto \varepsilon_2e^{i2\Phi^*_2},\;\; v_3e^{i3\Phi_3} \propto \varepsilon_3e^{i3\Phi^*_3}$. In contrast, the higher-order harmonics $v_4, v_5$ and $v_6$ arise from both the initial geometry and non-linear mixing of lower-order harmonics~\cite{Gardim:2011xv,Aad:2014fla}:
\begin{eqnarray}
\nonumber
&&v_4e^{i4\Phi_4} = a_{0}\; \varepsilon_4e^{i4\Phi^*_4}  + a_{1}\; \left(\varepsilon_2e^{i2\Phi_2^*}\right)^{2} +... =  c_0\;  e^{i4\Phi^*_4}  + c_{1}v_2^2e^{i4\Phi_2} +...\;,\\\nonumber
&&v_5e^{i5\Phi_5} = a_0 \varepsilon_5e^{i5\Phi^*_5} + a_1 \varepsilon_2e^{i2\Phi^*_2}\varepsilon_3e^{i3\Phi^*_3}+...= c_0  e^{i5\Phi^*_5} + c_1 v_2v_3e^{i(2\Phi_2+3\Phi_3)}+...\;,\\\label{eq:ep3a}
&&v_6e^{i6\Phi_6} = a_0 \varepsilon_6e^{i6\Phi^*_6} + a_1 \left(\varepsilon_2e^{i2\Phi^*_2}\right)^3+a_2 \left(\varepsilon_3e^{i3\Phi^*_3}\right)^2+a_3\varepsilon_2e^{i2\Phi^*_2}\varepsilon_4e^{i4\Phi^*_4}+...= c_0  e^{i6\Phi^*_6} + c_1 v_2^3e^{i6\Phi_2}+c_2 v_3^2e^{i6\Phi_3}+...\;\;\;\;\;
\end{eqnarray}
Furthermore, the measured $p(v_2)$ distributions show significant deviation in the tail from Bessel-Gaussian function\cite{Aad:2013xma}, suggesting either the $p(\varepsilon_2)$ is non-Gaussian~\cite{yan} or the response of $v_2$ to $\varepsilon_2$ is not strictly linear~\cite{niemi}. New ATLAS results show that the non-Gaussian behavior is captured by $<1-2\%$ difference between $v_2$\{4\} and $v_2$\{6\}, suggesting that $v_n\{2k\}$ for $k>1$ are not very sensitive to flow fluctuations given their current systematic uncertainties~\cite{dominik,Jia:2014jca}.

%In these methods, events in a narrow centrality interval are further classified according to the observed $v_m$ signal ($m=2$ and 3) in a forward rapidity range. This classification selects events with similar multiplicity but very different ellipticity or triangularity. The values of $v_n$ and EP correlations are then measured at mid-rapidity using the standard flow techniques.  
The EbyE flow fluctuations can also be studied using various event-shape selection methods~\cite{Schukraft:2012ah,Huo:2013qma}. The  performances of these methods have been validated in a AMPT model framework~\cite{Huo:2013qma,Jia:2014ysa} and applied in the ATLAS data analysis~\cite{ATLAS2014-022}. In this analysis, the azimuthal angle distribution of the transverse energy $\eT$ in the forward calorimeter over $3.3<|\eta|<4.9$ is expanded into a Fourier series for each event:
\vspace*{-0.1cm}
\begin{eqnarray}
\label{eq:res41}
2\pi\frac{d\eT}{d\phi} =  （\Sigma \eT）\left(1+2\Sigma_{n=1}^{\infty}q_n\cos n(\phi-\Psi_n)\right)
\end{eqnarray}
where the reduced flow vector $q_n$ represents the $\eT$-weighted raw flow coefficients $v_n^{\mathrm{obs}}$, $q_n=\Sigma \left(\eT v_n^{\mathrm{obs}}\right)/\Sigma \eT$.  Events are first divided according to $\Sigma \eT$ into centrality classes, events in each class are further divided according to the $q_2$. This classification separates events with similar multiplicity but with very different ellipticity. The values of $v_n$ are then calculated at mid-rapidity ($|\eta|<2.5$) using a 2PC method. 
%The non-flow effects are suppressed by a rapidity gap $|\Deta|>2$ between charged particle pairs.

\begin{figure}[!b]
\centering
\vspace*{-0.1cm}
\includegraphics[width=1\linewidth]{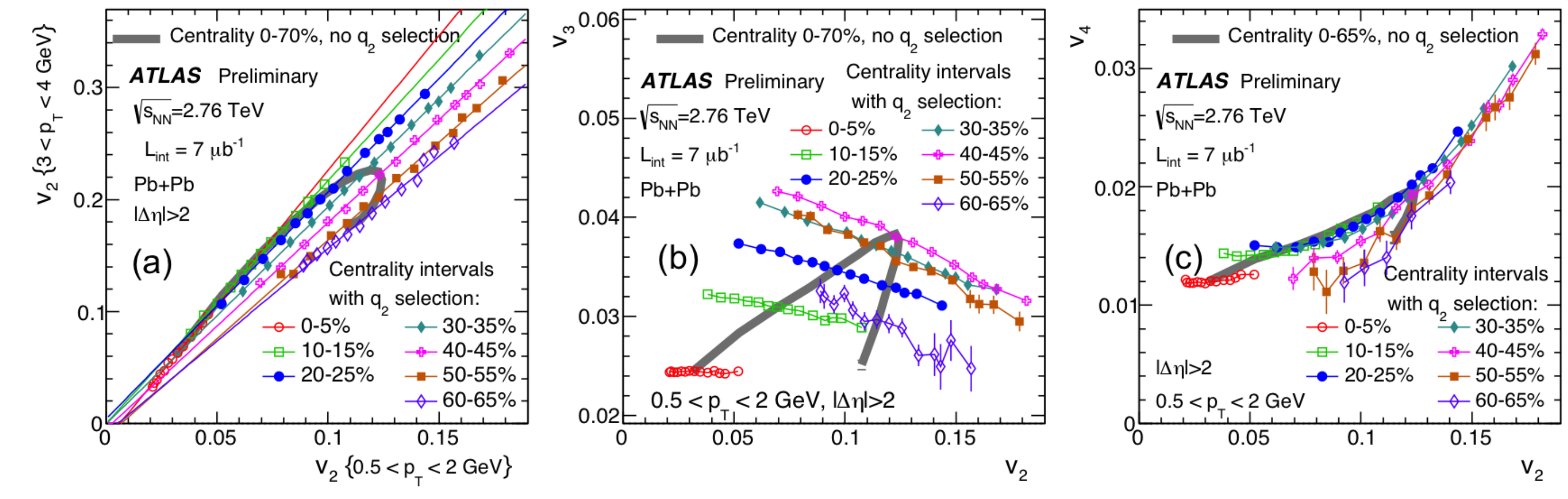}
\vspace*{-0.3cm}
\caption{\label{fig:4} The correlation of (panel a) $v_2$ in two different $\pT$ ranges, (panel b) $v_3$ and $v_2$ in the same $\pT$ range and (panel c) $v_4$ and $v_2$ in the same $\pT$ range. The data points in each centrality interval correspond to 14 different ellipticity classes selected via an event-shape engineering technique. These data are overlaid with the centrality dependence without event-shape selection (think grey lines). The thin solid straight lines in the left panel represent a linear fit of the data in each centrality. Results taken from Ref.~\cite{ATLAS2014-022}.}
\end{figure}

\begin{figure}[!t]
\centering
\includegraphics[width=0.9\linewidth]{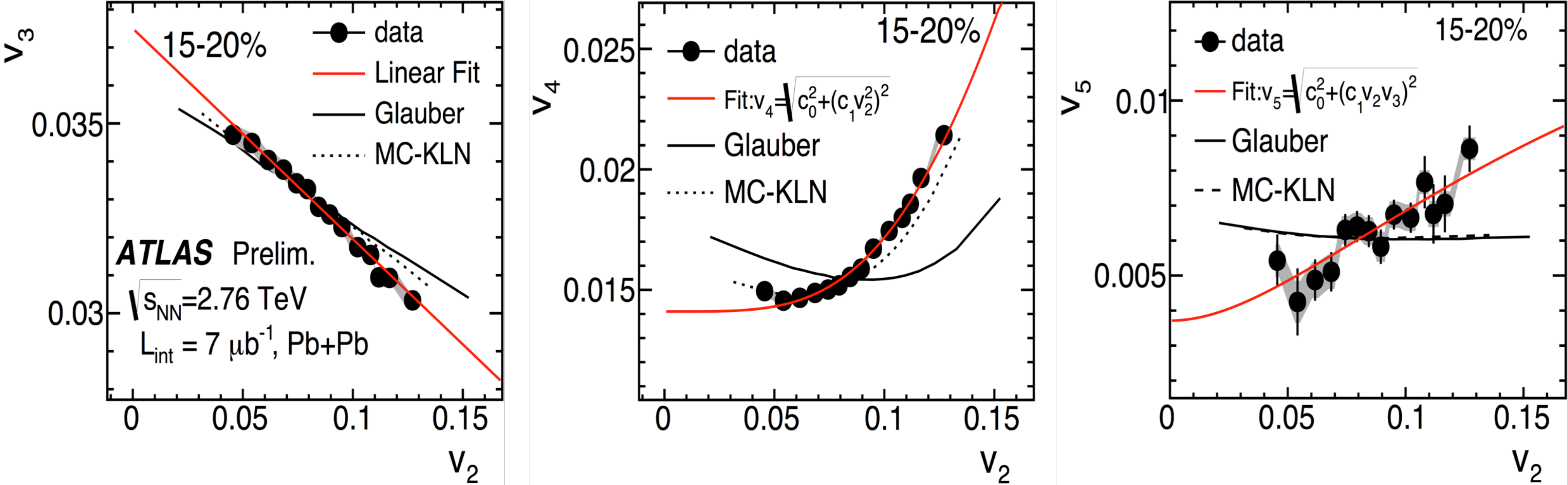}
\vspace*{-0.2cm}
\caption{\label{fig:5} The $v_3$--$v_2$ (left), $v_4$--$v_2$ (middle) and $v_5$--$v_2$ (right) correlations measured in $0.5<\pT<2$~GeV 10-15\% centrality interval. The correlation data are fit to functions that include both linear and non-linear contributions. The correlation data are also compared with re-scaled $\varepsilon_n$--$\varepsilon_2$ correlation from the MC Glauber and MC-KLN models. Results taken from Ref.~\cite{ATLAS2014-022}.}
\end{figure}

\begin{figure}[!t]
\centering
\includegraphics[width=0.41\linewidth]{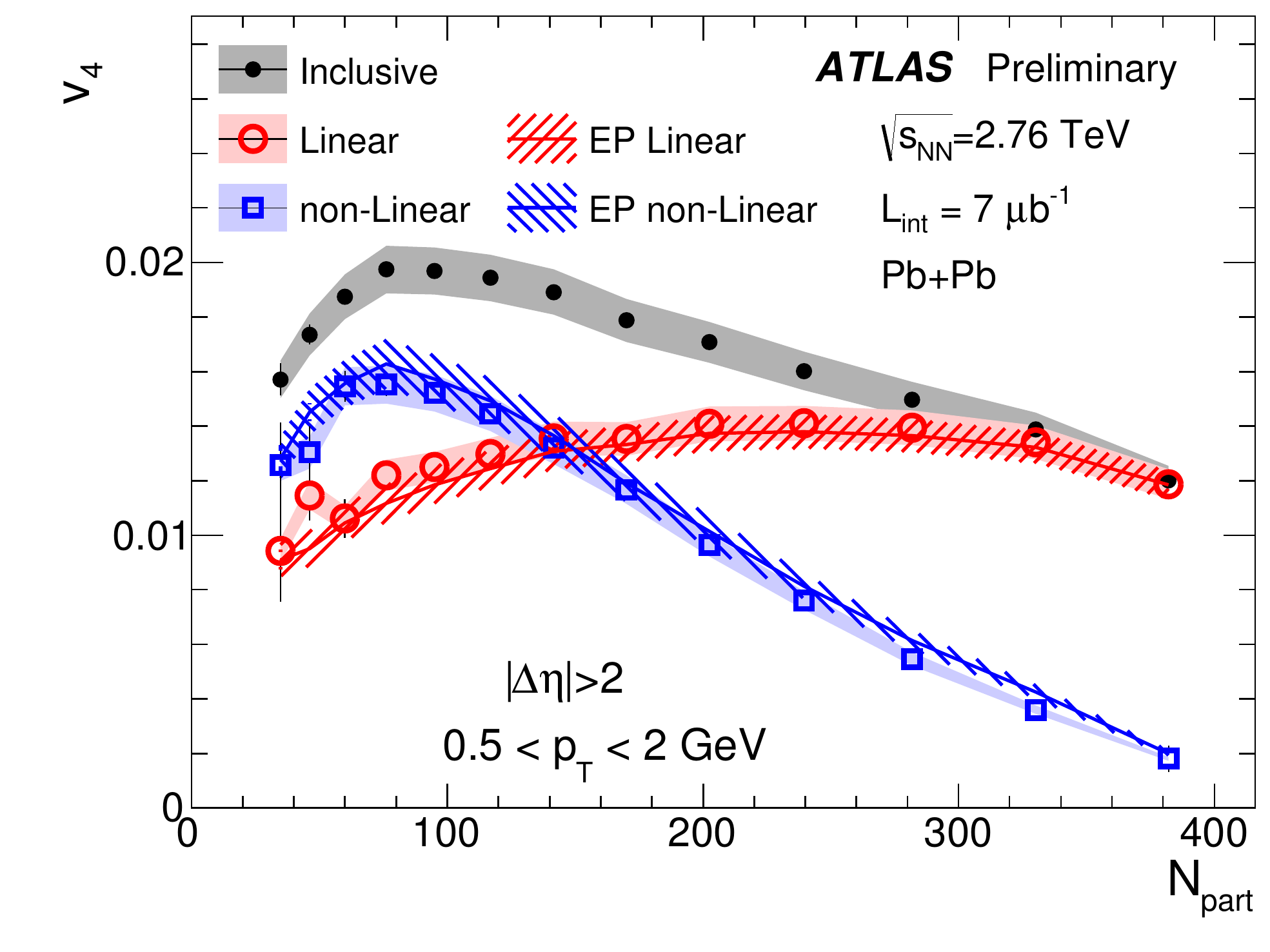}\includegraphics[width=0.41\linewidth]{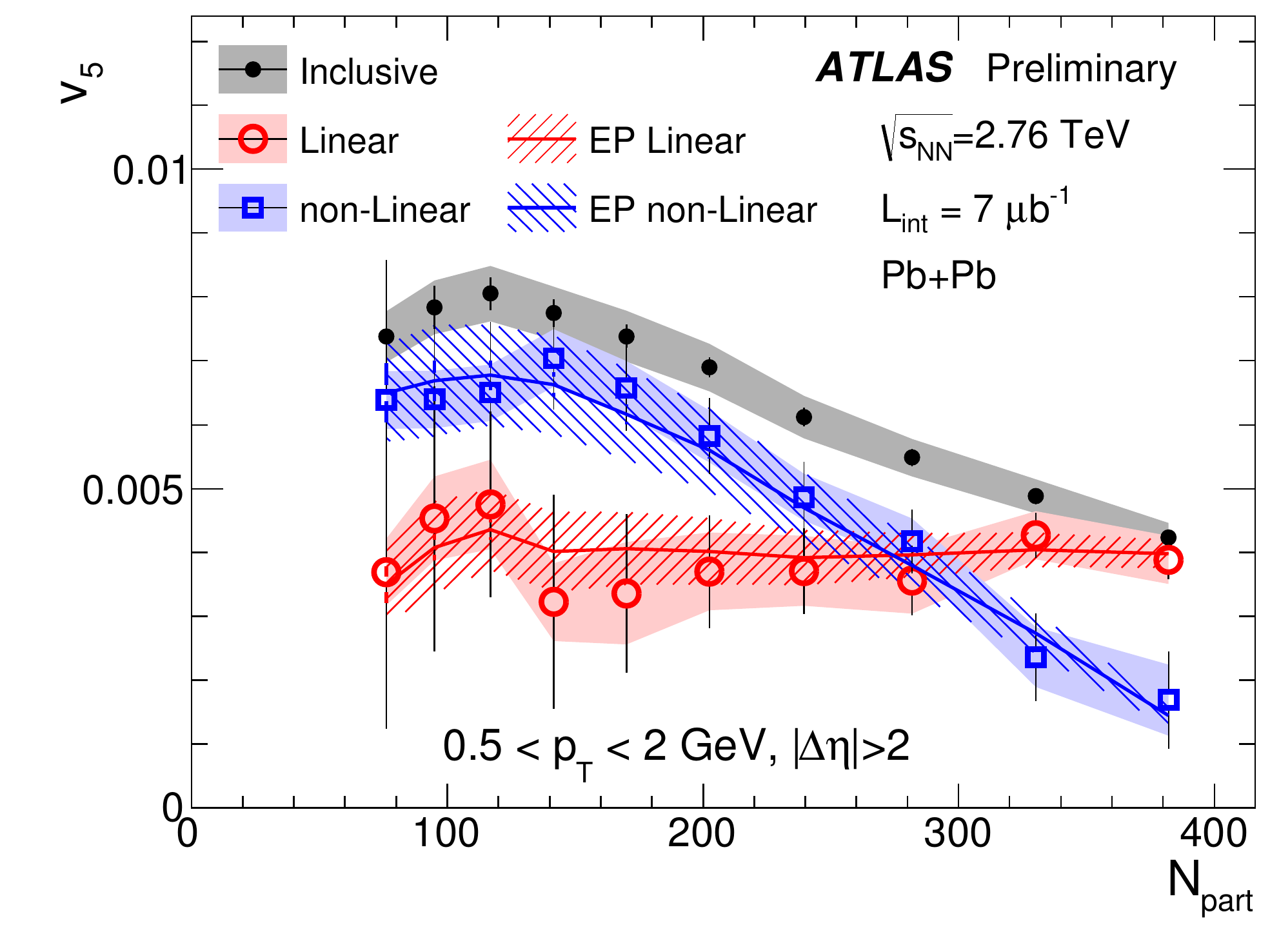}
\vspace*{-0.3cm}
\caption{\label{fig:6} The centrality dependence of the $v_4$ (left) and $v_5$ (right) in 0.5--2~GeV and the associated linear and non-linear components extracted from the fit in Figure~\ref{fig:5}. They are compared with the linear and non-linear components estimated from the previous published event-plane correlations~\cite{Aad:2014fla}. Results taken from Ref.~\cite{ATLAS2014-022}.}
\end{figure}

The correlation of $v_2$ between two different $\pT$ ranges from ATLAS, presented in Figure~\ref{fig:4}(a), shows a boomerang-like centrality dependence due to a stronger viscous correction for $v_2$ at higher $\pT$. In contrast, the correlation within a narrow centrality interval is found to be always linear,
suggesting that viscous effects are controlled by the system size not its overall shape. The $v_3$--$v_2$ correlations, shown in Figure~\ref{fig:4}(b), reveal an anti-correlation that is similar in magnitude to the correlation between $\varepsilon_3$ and $\varepsilon_2$ (see Figure~\ref{fig:5}(a)), implying that this correlation reflects mostly initial geometry effects. ATLAS also studied $v_4$--$v_2$ (Figure~\ref{fig:4}(c)) and $v_5$--$v_2$ correlations. The patterns in these correlations are found to be well described by two-parameter fits of the following form, motivated by interplay between the linear and non-linear collective dynamics given by Eq.~\ref{eq:ep3a}:
\begin{eqnarray}
\label{eq:res}
v_4= \sqrt{c_{0}^2 + (c_{1}v_2^2)^{2}}\;, v_5= \sqrt{c_{0}^2 + (c_{1}v_2v_3)^{2}}\;,
\end{eqnarray}
These fits allow ATLAS to decompose the $v_4$ and $v_5$, centrality by centrality, into linear and non-linear terms as $v_n^{\rm{L}}=c_0$ and $v_n^{\rm{NL}}=\sqrt{v_n^2-c_0^2}$, respectively, as shown in Figure~\ref{fig:6}. The linear term associated with $\varepsilon_n$ depends only weakly on centrality, and dominates the $v_n$ signal in central collisions. The excellent agreement with a similar decomposition based on the measured EP correlations~\cite{Aad:2014fla} implies that the correlations between flow magnitudes arise mostly from the correlations between the flow angles.

\vspace*{-0.3cm}
\section{Future prospects}
The application of event-shape selection techinique is not only limited to flow observables. It should be straightforward to apply this technique to other observables, such as HBT correlation, $R_{\mathrm{AA}}$, dihadron correlation, chiral magnetic effects and more. A study by PHENIX~\cite{esumi} shows that the measured 2$^{\mathrm{nd}}$-order freezeout eccentricity $\varepsilon_2^{\rm f}$ is strongly correlated with selection on the ellipticity. This is qualitatively expected, as events with large $v_2$ should on average have large $\varepsilon_2$, and hence they may have larger $\varepsilon_2^{\rm f}$.

\begin{figure}[!h]
\vspace*{-0.2cm}
\includegraphics[width=1\linewidth]{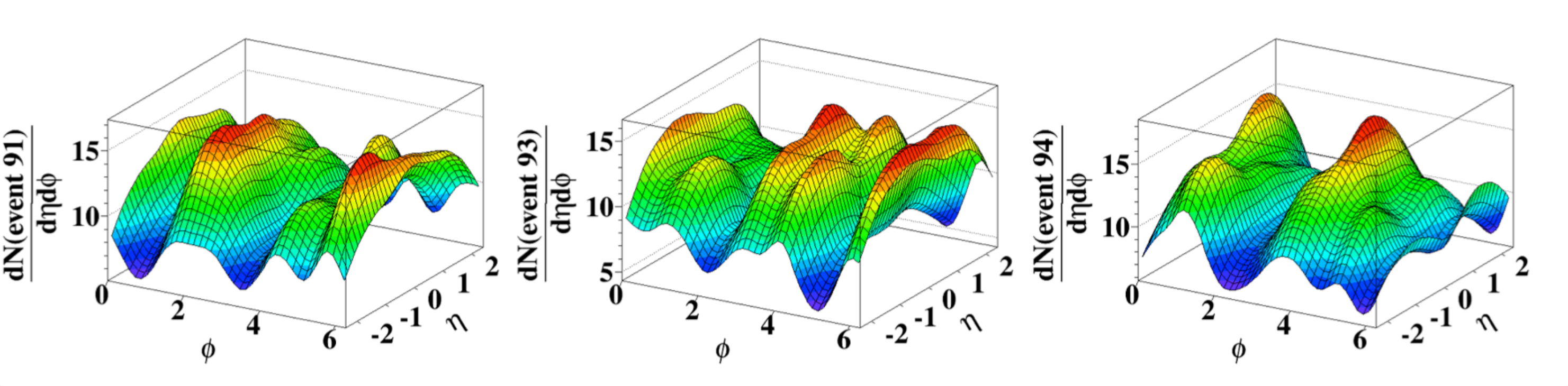}
\vspace*{-0.2cm}
\caption{\label{fig:8} The distributions of the particles in $\eta\times\phi$ space for three typical Au+Au collisions at $\sqrtsnn=200$ GeV in 0--10\% centrality interval, obtained from 3+1D calculation based on ideal hydrodynamics~\cite{pang}.}
\end{figure}

One open issue is the response or back-reaction of the medium to the energy deposited by jets. CMS results suggest that the lost energy of very high $\pT$ jets are transported to very large angle and to low $\pT$ particles~\cite{Chatrchyan:2011sx}, but not much is know about the back-reaction of the mini-jets that have the energy of a few GeV to few ten's of GeV. These mini-jets are abundantly produced and dominate particle spectrum at intermediate $\pT$. The back-reaction of mini-jets was historically studied using 2PC analysis aided by flow background subtraction, which suffers large systematics due to dominance of collective flow in the correlation structure. We can improve the situation by performing the analysis in events selected to have small $v_n$. But ultimately, 2PC method may not be the best approach to investigate the jet-medium interactions. Instead, we need methods that looks directly the localized $\eta\times\phi$ structures in the EbyE particle multiplicity or $\eT$ distributions. To motivate this idea, Figure~\ref{fig:8} shows the EbyE $\eta\times\phi$ distribution of particle density from a 3+1D hydrodynamic calculation based on the AMPT initial condition~\cite{pang}. The localized peaks and valleys in these events could be remnant of the mini-jets in the initial state. These localized structures, or ``hydro-jets'', are much broader than typical high-$\pT$ jets. They can be found, possibly as fake-jets, by running standard jet reconstruction algorithm. Obviously, the jet finding algorithm may need to be modified in order to maximize the finding efficiency and better adapt to the shape of these objects. One can then perform a detailed study of the spectrum and substructure of these hydro-jets.

\begin{figure}[!h]
\centering
\vspace*{-0.2cm}
\includegraphics[width=0.8\linewidth]{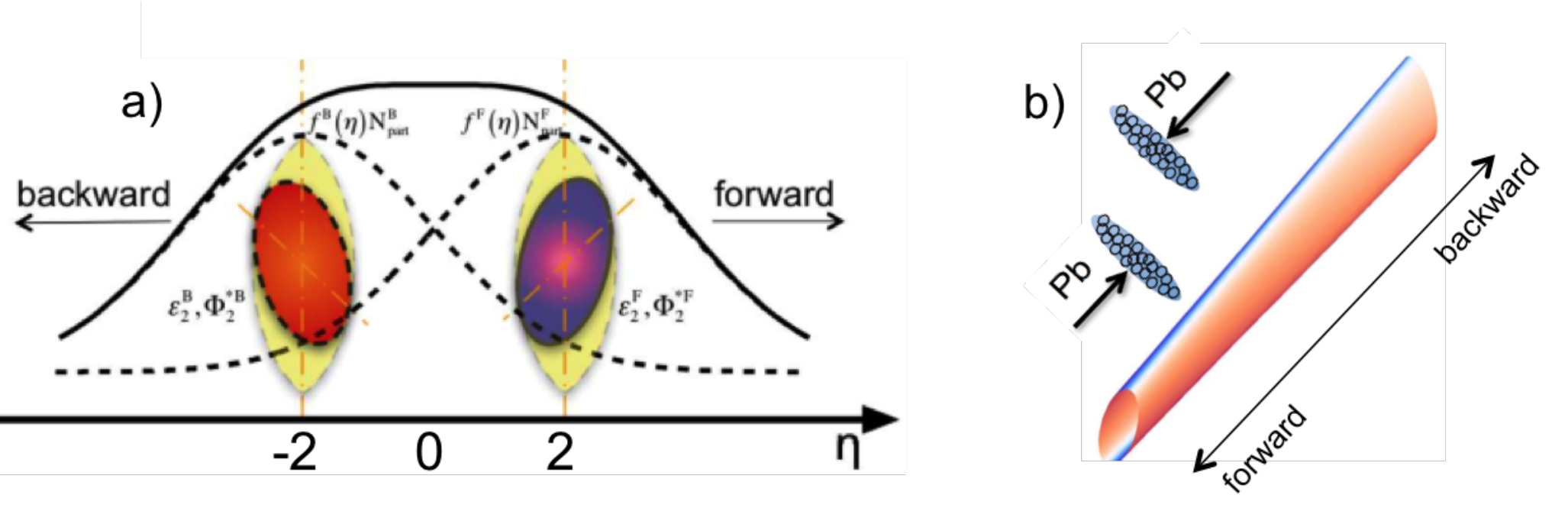}
\vspace*{-0.3cm}
\caption{\label{fig:9}  Schematic illustration of the forward-backward fluctuation of second-order eccentricity and participant plane in A+A collisions. The dashed-lines indicate the particle production profiles for forward-going and backward-going participants, respectively~\cite{Jia:2014ysa,Bozek:2010vz}.}
\end{figure}

Most studies in A+A collisions consider only fluctuations in the transverse plane, and dynamics in the longitudinal direction are often assumed to be boost invariant. However, a Glauber model calculation~\cite{Jia:2014ysa} shows that the number of participating nucleons and eccentricity vectors $\vec{\varepsilon}_n\equiv(\varepsilon_n,\Phi_n*)$, defined separately for the two colliding nuclei, can differ strongly on EbyE basis due to fluctuations: $\npartf\neq\npartb$ and $\vec{\varepsilon}_n^{\;\mathrm{F}}\neq \vec{\varepsilon}_n^{\;\mathrm{B}}$. Furthermore, the energy deposition for each participating nucleon is not symmetric: Particles in the forward (backward) rapidity are preferably produced by the participants in the forward-going (backward-going) nucleus~\cite{Bialas:2010zb}. Due to these two effects, the eccentricity vector and the flow vector $\vec{v}_n\equiv (v_n,\Phi_n)$ of the initially produced fireball should depend on $\eta$ (see Figure~\ref{fig:9})~\cite{Jia:2014ysa}:
\begin{eqnarray}
\label{eq:pros1}
\vec{v}_n(\eta) \propto \vec{\varepsilon}^{\mathrm{\;tot}}_n(\eta)\approx \alpha(\eta)\vec{\varepsilon}_n^{\;\rm F}+(1-\alpha(\eta))\vec{\varepsilon}_n^{\;\rm B}\equiv\varepsilon^{\mathrm{tot}}_n(\eta)e^{in\Phi_n^{*\rm tot}(\eta)}, \alpha(\eta) \approx\frac{ f(\eta) }{f(\eta)+f(-\eta)}\;, n= 2 \;\mathrm{or}\; 3
\end{eqnarray} 
where $\alpha(\eta)$ is a $\eta$-dependent weight factor for the forward-going participating nucleons, and $f(\eta)$ is the average emission profile per-nucleon~\cite{Bialas:2010zb}. This relation has been verified using the Glauber model and AMPT model. The asymmetry between $\vec{\varepsilon}_n^{\;\rm F}$ and $\vec{\varepsilon}_n^{\;\rm B}$ are very large and they are shown to be converted into similar asymmetry of $\vec{v}_n$ between forward and backward $\eta$. This F/B asymmetry is expected to be particularly large for $\pPb$ collisions, and initial indication of such F/B asymmetry has indeed been observed in $\pPb$ collisions~\cite{xu}.

\vspace*{-0.3cm}
\section{Summary}
The Quark Matter 2014 conference has seen impressive progresses in the study of collective behavior in high energy nuclear collisions. Significant azimuthal harmonics have been measured for $v_1$ to $v_5$ in $\pA$ collisions over broad range of $\pT$ and event multiplicity. The $v_2$ signal from multi-particle correlations are large, suggesting that the ridge in $\pA$ collisions reflects the genuine collectivity of the produced medium. Remarkable agreement of the $v_n(\pT,N_{\mathrm{ch}})$ between $\pA$ and A+A collisions suggests that the observed collectivity is driven by similar dynamics in the two systems. Several new flow results in A+A collisions, such as $v_n$ for different particle species and collision systems and $v_n$ in ultra-central collisions, serve to refine the hydrodynamic models. Important progresses have also been made in measuring a large class of event-by-event flow observables, in particular a brand-new detailed study of the $v_2$--$v_2$ and $v_n$--$v_2$ correlations using the event-shape selection technique. A coherent picture of initial condition and collective flow based on linear and non-linear hydrodynamic responses is derived, which qualitatively describe most results on these observables. Several new types of fluctuation measurements should be pursued, which can further our understanding of the event-shape fluctuations and collective expansion dynamics.

This research is supported by NSF under grant number PHY-1305037 and by DOE through BNL under grant number DE-AC02-98CH10886.
%% \begin{figure}
%% \begin{center}
%% \includegraphics*[width=9.cm]{bessel1}
%% \caption{}
%% \label{fig:}
%% \end{center}
%% \end{figure}

%% The Appendices part is started with the command \appendix;
%% appendix sections are then done as normal sections
%% \appendix

%% \section{}
%% \label{}

%% References
%%
%% Following citation commands can be used in the body text:
%% Usage of \cite is as follows:
%%   \cite{key}         ==>>  [#]
%%   \cite[chap. 2]{key} ==>> [#, chap. 2]
%%

%% References with BibTeX database:

%\bibliographystyle{elsarticle-num}
%\bibliography{<your-bib-database>}

%% Authors are advised to use a BibTeX database file for their reference list.
%% The provided style file elsarticle-num.bst formats references in the required Procedia style

%% For references without a BibTeX database:

%\bibliographystyle{atlasBibStyleWoTitle}
%\bibliography{QM14_Jiangyong_Jia_v2}

\end{document}